# Gas Sensing with *h*-BN Capped MoS₂ Heterostructure Thin Film Transistors

G. Liu, S.L. Rumyantsev, C. Jiang, M.S. Shur and A.A. Balandin

*Abstract* — **We have demonstrated selective gas sensing with molybdenum disulfide (MoS₂) thin films transistors capped with a thin layer of hexagonal boron nitride (*h*-BN). The resistance change was used as a sensing parameter to detect chemical vapors such as ethanol, acetonitrile, toluene, chloroform and methanol. It was found that *h*-BN dielectric passivation layer does not prevent gas detection via changes in the source-drain current in the active MoS₂ thin film channel. The use of *h*-BN cap layers (thickness *H*~10 nm) in the design of MoS₂ thin film gas sensors improves device stability and prevents device degradation due to environmental and chemical exposure. The obtained results are important for applications of van der Waals materials in chemical and biological sensing.**

## I. INTRODUCTION

GAS sensing technology based on the relative resistance change upon the gas molecule adsorption and desorption enables fast speed and low cost sensors. Two-dimensional (2D) materials – also referred to as *van der Waals materials* – such as graphene and MoS₂ are natural candidates for gas sensing applications owing to the ultimately high surface – to – volume - ratio of the 2D materials and the wide-range tunable Femi level [1-8]. The molecule adsorption and desorption on the surface of 2D channels tunes the local Fermi level and changes the resistance of the 2D channels. Unlike the zero band-gap graphene MoS₂ has a sizable energy band gap, which ranges from ~1.1 to 1.9 eV [6-8] for the bulk and monolayer MoS₂ samples, respectively. The wide band gap of MoS₂ thin films results in a stronger effect on the source – drain current

The work of A. A. Balandin and G. Liu was supported in part by the National Science Foundation (NSF) through the grants CCF-1217382 and CMMI-1404967; by the Semiconductor Research Corporation (SRC) and by the Defense Advanced Research Project Agency (DARPA) within the Semiconductor Technology Advanced Research Network through the Center for Function Accelerated nanoMaterial Engineering (FAME). The work of S. L. Rumyantsev and M. S. Shur was supported by NSF through the EAGER Program. SLR acknowledges also partial support from RFBR.

A. A. Balandin and G. Liu are with the Nano-Device Laboratory (NDL) and Phonon Optimized Engineered Materials (POEM) Center, Department of Electrical and Computer Engineering, Bourns College of Engineering, University of California – Riverside, Riverside, California 92521 USA (e-mail: balandin@ece.ucr.edu).

S. L. Rumyantsev and M. S. Shur are with the Department of Electrical, Computer, and Systems Engineering, Center for Integrated Electronics, Rensselaer Polytechnic Institute, Troy, New York 12180, USA

S. L. Rumyantsev is also with the Ioffe Physical-Technical Institute, St. Petersburg 194021, Russia

produced by the analyte molecules attached to the surface. We have recently demonstrated experimentally that the relative resistance change in MoS₂ thin film transistors (TFTs) is better sensing parameter than that in graphene devices [5]. However, the graphene sensors are more suited for using the low-frequency current fluctuations as an additional sensing parameter [2, 4, 9].

An important question, which arises in the context of the sensor applications of all van der Waals materials, is whether it is possible to use any protective cap layer or surface passivation without degrading the 2D channel sensing ability. A prolonged exposure of TFTs with channels made of MoS₂ and other 2D materials degrades their sensing performance owing to oxidation and surface contamination [10-11]. In this Letter, we show that the *h*-BN capped MoS₂ heterostructure thin film transistors (HTFTs) retain their ability for gas sensing while becoming more robust and resistive to degradation.

## II. DEVICE FABRICATION

For the proof-of-concept demonstration, the thin films of MoS₂ were mechanically exfoliated from the bulk crystal and transfer onto Si/SiO₂ substrate (300-nm thick oxide). The thickness of the resulting films was in the range from 2 nm to 9 nm as confirmed by the atomic force microscopy (AFM) measurements. The capping process was carried out right after identifying the desired MoS₂ layer for the channel fabrication. The selected *h*-BN layers were exfoliated and transferred onto the polydimethylsiloxane (PDMS) stamp. The stamp was then attached to a glass substrate. After the thin *h*-BN layer was identified by an optical microscope, the entire *h*-BN/PDMS/glass stack was mounted on an alignment stage. The stage movement was controlled by a micro-manipulator to accurately position the capping onto the target MoS₂ film [12]. We intentionally used the *h*-BN layers smaller in lateral size than the target MoS₂ channel layers so that the uncapped regions can be used to make contacts to metal electrodes. After the *h*-BN layer was placed on top of MoS₂ film, the PDMS stamp was peeled off leaving behind the *h*-BN capped MoS₂ thin-film channel. A standard electron-beam lithography (EBL) was used to generate a pattern for the source and drain contacts. The contacts were made by evaporating 60 nm of gold. The device schematics and optical images are shown in Figure 1. In order to accurately compare the capped and uncapped devices we fabricated both types of structures on the same MoS₂ layers.

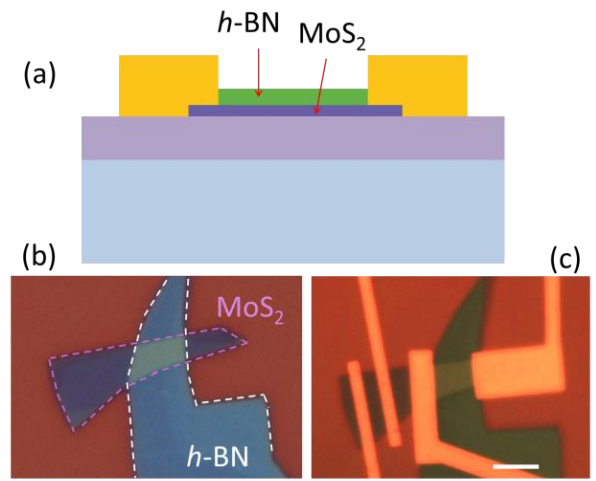

Figure 1: (a) Schematics of the *h*-BN capped MoS₂ TFT. (b and c) Optical images of a representative *h*-BN capped MoS₂ stack and sensor device. The dash lines outline the *h*-BN and MoS₂ layers. The scale bar is 5 μm.

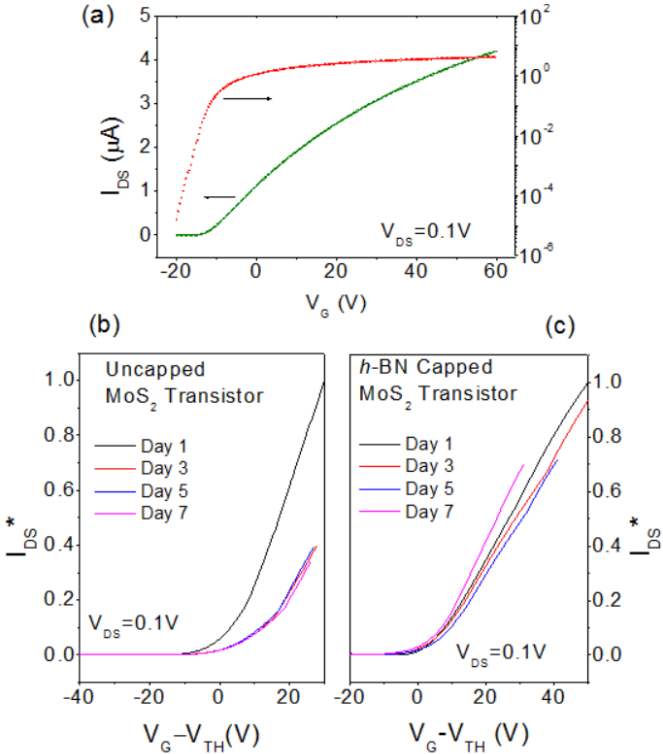

Figure 2: (a) Typical transfer characteristic of *h*-BN capped MoS₂ TFT. The mobility is around 30-50 cm²/Vs. The on-off ratio is on the order of 10⁵-10⁶. The ambient aging comparison of the uncapped and capped device is shown in (b) and (c), respectively. The characteristics of the uncapped device MoS₂ degraded 60% after two days of aging while the capped device did not degrade much within a week. Note that the high gate bias is due to back gating via 300-nm thick SiO₂ layer in these prototype devices.

The typical transfer characteristics of *h*-BN capped thin film MoS₂ device are shown in Figure 2 (a). The mobility of our devices is in the range of 30-50 cm²/Vs as determined from the formula $\mu = \frac{L}{W} \frac{1}{C_G V_{DS}} \frac{\Delta I_{DS}}{\Delta V_G}$, where $L$ and $W$ are the length and width of the channel and $C_G$ is the capacitance per unit area.

The on-off ratios were extracted in the range of 10⁵-10⁶.

Under ambient conditions the uncapped devices started to degrade in the first couple of days as seen in Figure 2 (b). The source-drain current in transfer *I-V* characteristics decreased by 60% after two days of ambient aging and then stabilized in the following days. To clearly illustrate the aging effect, we normalized the source-drain current $I_{DS}$ to the maximum value ($I_{DS}$ =1.34 μA at $V_G$=50 V of the first day of measurements). We also noticed a minor threshold voltage $V_{TH}$ shift as the ambient environment doped the device p-type. To compensate for the doping effect we plotted $V_G$ - $V_{TH}$ on the *X*-axis.

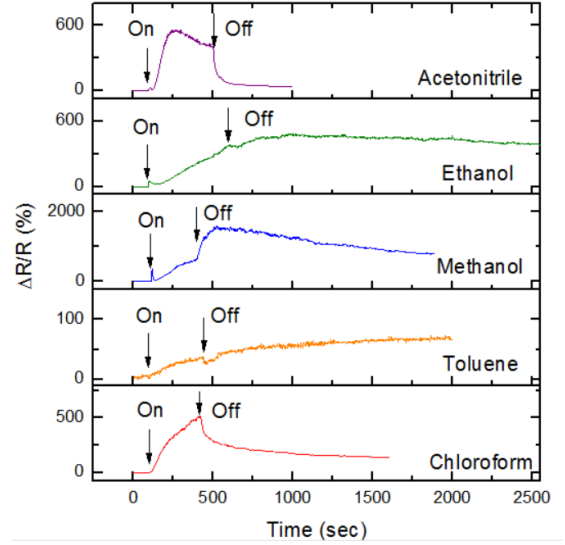

Figure 3: Response of *h*-BN capped MoS₂ sensor to different gas vapor. The thickness of MoS₂ is 9 nm. The current is monitored by applying $V_{DS}$=0.1 V.

The *h*-BN capped MoS₂ HTFTs were more robust under ambient condition as can be seen in Figure 2 (c). The transfer characteristics did not degrade substantially over a week period of time. The I-V curves have been plotted in the same way as for the uncapped devices in Figure 2 (b). The current is normalized to its maximum value ($I_{DS}$ =0.96 μA at $V_G$=50 V of the first day of measurement). Note that the aspect ratio *L/W* of the uncapped and capped devices is 0.22 and 1.5, respectively. It is important to note that the capped devices had much larger on-current density than the uncapped ones. Since *h*-BN capping was introduced before EBL process the MoS₂ channel layer was free from e-beam resist contamination. The enhanced performance is therefore might reflect the fact that *h*-BN MoS₂ TFTs are residue free.

## III. GAS SENSING RESULTS

To test *h*-BN-MoS₂ HTFT senor operation, the vapors were produced by bubbling dry air through the respective solvents and diluting the gas flow with the dry air. The resulting concentrations were ~0.5 *P/Pₒ*, where *P* was the vapor pressure and *Pₒ* was the saturated vapor pressure. When the device was exposed to the vapors, the vapor molecules were attaching to the channel surface, thus, creating negative or positive surface charges at the *h*-BN capped MoS₂ channel. The molecules introduced n-type or p-type doping effect depending on the vapor species. We selected the same solvents in order to compare with our previous finding of gas sensing on uncapped MoS$_2$ TF-FETs [5]. Specifically, we used polar solvent: acetonitrile (CH$_3$CN), ethanol (C$_2$H$_5$OH), methanol (CH$_3$OH), and non-polar solvents: toluene (C$_6$H$_5$-CH$_3$), chloroform (CHCl$_3$).

The source-drain current was monitored as $h$-BN-MoS$_2$

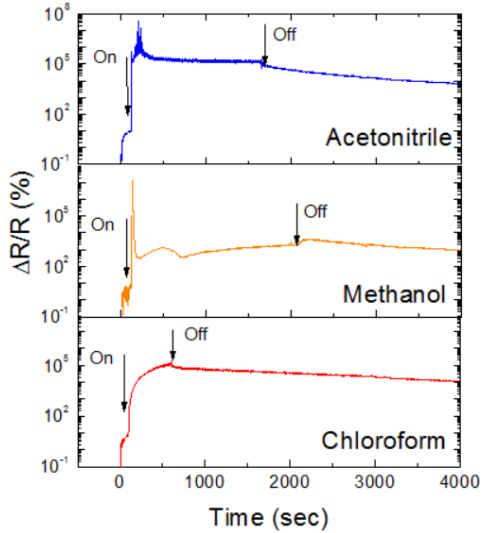

Figure 4: Response of $h$-BN capped MoS$_2$ HFT different vapors. The thickness of MoS$_2$ channel is 2 nm.

HTFT senor was exposed to different gases. For all the measurements, we kept $V_{DS}$=0.1 V and $V_G$=0 V. Figure 3 shows the sensitivity ($\Delta R/R$) as the gases turned on and off. For all the polar solvents, the resistance sharply increased after the gas was turned on. However, as the gas was turned off, the behavior quite different. For acetonitrile, the resistance restored to the initial value; whereas for methanol and ethanol, the resistance first increased, and after certain time began to restore at a slow speed. For the two non-polar solvents, the responses also diverged. Under the exposure to chloroform, the resistance kept increasing until the gas was turned off. The resistance partially restored to the initial value. When exposed to toluene, the resistance response was much weaker and slower. As toluene gas was turned off the resistance continued to increase expect for a short time of kink. The resistance did not restore after a long waiting time. The measurements were repeated after several days. The results were reproducible.

We also checked the gas sensing with thinner MoS$_2$ devices ($H$=2 nm). Owing to even higher surface-to-volume ratio the $\Delta R/R$ in the 2-nm thick channel HTFT upon exposure to gas vapors is 10$^2$-10$^3$ higher than that in the 9-nm thick devices. However, the resistance restoration was much slower than in the 9-nm device.

The responses of $h$-BN-MoS$_2$ HTFTs on gases were different than those of the uncapped MoS$_2$ TFTs [5]. In the uncapped MoS$_2$ TFTs devices the resistance decreased when exposed to polar solvent and increased when exposed to non-polar solvent. However, the important result is that the capping MoS$_2$ TFTs with h-BN layer does not prevent the gas detection while simultaneously improving device stability and resistance to environmental degradation. This is in contrast to the Al$_2$O$_3$ passivation of MoS$_2$ channels which eliminated the sensor action [5]. The differences could be related to differences in the dielectric constants of the materials and the fact that exfoliated $h$-BN layer has atomically sharp interface with the MoS$_2$ channel.

## IV. CONCLUSION

This work demonstrates that $h$-BN dielectric passivation layer preserves the ability of MoS$_2$ TFTs for sensitive and selective gas detection. In addition, the use of $h$-BN cap layers in MoS$_2$ thin film gas sensors improves the stability of the devices and prevents their degradation due to environmental exposure.

## REFERENCES

[1] F. Schedin *et al.*, "Detection of individual gas molecules adsorbed on graphene," *Nature Mater.*, vol. 6, no. 9, pp. 652-655, Jul. 2007.

[2] S. Rumyantsev, G. Liu, M. Shur, R.A. Potyrailo and A.A. Balandin, "Selective gas sensing with a single pristine graphene transistor," *Nano Lett.*, vol. 12, no. 5, pp. 2294-2298, Apr. 2012.

[3] F. K. Perkins *et al.*, "Chemical vapor sensing with monolayer MoS$_2$," *Nano Lett.*, vol. 13, no. 2, pp. 668-673, Jan. 2013.

[4] S. Rumyantsev, G. Liu, R.A. Potyrailo, M.S. Shur and A.A. Balandin, "Selective sensing of individual gases using graphene devices," *IEEE Sensors J.*, vol. 13, no. 8, pp. 2818-2822, Mar. 2013.

[5] R. Samnakay, C. Jiang, S. L. Rumyantsev, M.S. Shur, and A.A. Balandin, "Selective chemical vapor sensing with few-layer MoS$_2$ thin-film transistors: comparison with graphene devices," *App. Phys. Lett.*, vol. 106, no. 2, pp. 023115-1-023115-5, Jan. 2015.

[6] R. Ganatra and Q. Zhang, "Few-Layer MoS$_2$: a promising layered semiconductor," *ACS Nano*, vol. 8, no. 5, pp. 4074-4099, Mar. 2014.

[7] K. F. Mak, C. Lee, J. Hone, J. Shan, and T. F. Heinz, "Atomically thin MoS$_2$: A new direct-gap semiconductor," *Phys. Rev. Lett.*, vol. 105, No. 13, pp. 136805-1-136805-4, Sep. 2010.

[8] D. J. Late *et al.*, "Sensing Behavior of Atomically Thin-Layered MoS2 Transistors," *ACS Nano*, vol.7, no. 6, pp 4879–4891, May 2013.

[9] S. L. Rumyantsev, C. Jiang, R. Samnakay, M. S. Shur, and A. Balandin, "1/$f$ Noise Characteristics of MoS$_2$ Thin-Film Transistors: Comparison of Single and Multilayer Structures," *IEEE Electron Device Lett.*, vol. 36, no. 5, pp. 517-519, May 2015.

[10] W. Park *et al.*, "Oxygen environmental and passivation effects on molybdenum disulfide field effect transistors," *Nanotechnology*, vol. 24, no. 9, pp. 095202-1-095202-5, Feb. 2013

[11] S. KC, R. C. Longo, R. M. Wallace, and K. Cho. "Surface oxidation energetics and kinetics on MoS$_2$ monolayer," *J. Appl. Phys.*, vol. 117, no.13 pp. 135301-1-135301-9, Apr. 2015.

[12] C.H. Lee *et al.*, "Atomically thin p–n junctions with van der Waals heterointerfaces," Nature Nanotech., vol. 9, no. 9, pp. 676-681, Aug. 2014.